# Google Web APIs – an Instrument for Webometric Analyses?

Philipp Mayr[*], Fabio Tosques[**]

[*]*mayr@bonn.iz-soz.de*
Informationszentrum Sozialwissenschaften (Social Science Information Centre), Department Research and Development, Lennéstr. 30, D-53113 Bonn (Germany)

[**]*abutux@web.de*
Institute of Library Science, Humboldt University, D-Berlin (Germany)

**Introduction**
The analysis of the Internet and its applications touches upon varied fields of research such as information science, computer science, economics, and psychology. One reason for the growing scientific interest in the Internet is the already high and still growing amount of web users, applications, contents, and web servers worldwide. Despite a variety of different search engines, there is no doubt about the predominance of Google search engine technology today. Among the factors that have added decisively to Google's success are that over a long period of time, Google has offered the largest index, innovative new services and highly optimized performance and usability.

This poster introduces Google Web APIs (Google APIs) as an instrument and playground for webometric studies. Several examples of Google APIs implementations are given. Our examples show that this Google Web Service can be used successfully for informetric Internet based studies albeit with some restrictions. For instance, we can show that hit results from the two different Google interfaces: Google APIs and the standard interface Google.com (Google Standard) vary in range, structure und availability. Our poster demonstrates first results of our research with Google APIs, gives implementation examples and makes possibilities and restrictions of the Google APIs clearer.

**Google APIs**
In spring 2002, Google decided to allow Internet researcher automated queries for the first time and published the interfaces Google Web APIs available at http://www.google.com/apis. The APIs are implemented as a web service that supports different SOAP (Simple Object Access Protocol) methods which are described in a WSDL (Web Services Description Language) file. Web Services are supported by all common programming languages among them Java, .NET (VBA, C#), C++, Python, PHP etc. To implement our queries, we used the language Perl.

We have been testing the Google APIs as a scientific tool for web data gathering since last year. In scientific publications, Google APIs have until now only been mentioned in passing (Thelwall, 2004). In the information science scholarly discourse, no intensive research on Google APIs has been published yet. To close this gap, we want to exemplarily introduce Google APIs. In order to secure comparability of Google APIs and the Google Standard search, hit results of both interfaces are juxtaposed.

*Analyses with the APIs*
The analysis of hit results of the big search engines is considered as a standard tool in webometrics or cybermetrics already for several years. Various webometric studies gained attention within the scientific community (Almind & Ingwersen, 1997, Bar-Ilan, 1998, Bar-Ilan, 2002, Rousseau, 1997, Rousseau, 1998). Our following analyses are closely connected to these webometric papers and try to transfer their concepts to the Google APIs. See also our project page where we set up some demo Google APIs implementations available at http://bsd119.ib.hu-berlin.de/~ft/index_e.html.

We performed the following analyses and hope to stimulate informetric researchers to evaluate and use the, so far, unused Google APIs for their Internet research.

1. The analysis of time series: Analysis of a set of standard queries beginning July 2004 and comparison with Google.com results (see cutout in Figure 1 below).
2. Journal web coverage: Testing coverage of the complete ISI journal list (more than 11,000 journal titles) on the web performing hit counts and backlink analysis.
3. Top level domain (TLD) analysis: Transfer the idea of Rousseau (Rousseau, 1997) to the Google APIs and found the Lotka function in TLD distributions. Our APIs demos enable live TLD analysis (see Figure 2 below).
4. Distribution of file formats on the web: An application which visualizes the distribution of file formats on the Google Web for an entered query.

**Results**
The analysed data in the time series show very clearly that for all our repeated queries Google Standard searches in a much larger and different index than Google APIs. Figure 1 displays the hit results for the query *webometrics* from both interfaces in a time series. Obviously Google Standard comes up with more results and larger fluctuations in the data. A more detailed observation of the APIs results show Google Standards ups and downs some days later (APIs = smaller and less updated index).

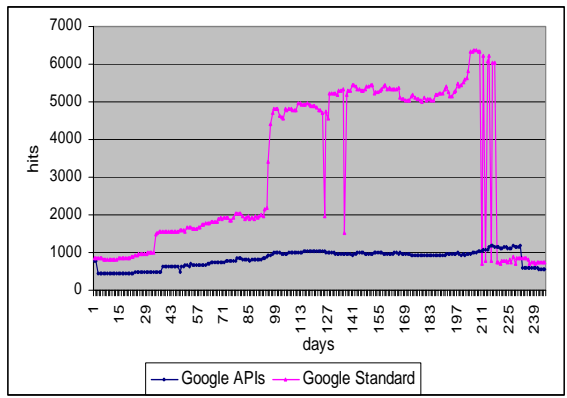

Figure 1: The query *webometrics*. Hit results in Google Standard and Google APIs.

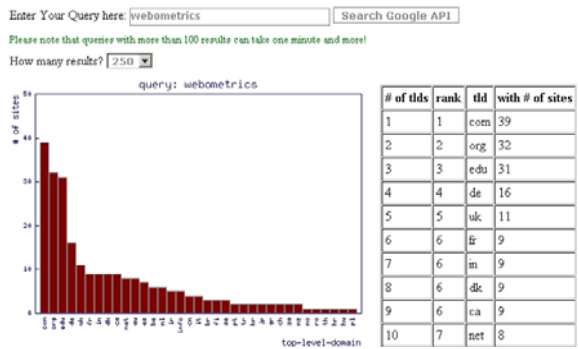

Figure 2: Screenshot of a power law TLD distribution of 250 URLs for the query *webometrics* via a live analysis with Google APIs.

The difference in the underlying data of the two search interfaces is more noticeable by comparing specific hit lists. Document ranking is therefore not the same. Surprisingly, there are sometimes more differing URLs in the result sets (100 hits per query were analysed) returned by Google APIs. One possible explanation for these result differences could be optimization in Google Standard.

Regarding the reliability of Google APIs, it was observed that the service did not function in the same way at all times. In view of service availability and performance, Google APIs has some disadvantages (for example, performance for more than 100 hits).

**Conclusions**

The Web service Google APIs, which weirdly continues to hold its beta status, is an interesting subject for webometric research. As there are few obstacles in the use of the Google APIs, an implementation of one's own analyses can be reached quite quickly. First of all it has to be clear that querying the Google APIs does not deliver the same result data as the highly optimized Google Standard interface. So far we can recommend the Google Web APIs for data generating and processing at least for prototyping purposes. Despite individual restrictions (only 10,000 APIs hits per day) and problems, the positive aspects of the, by now, elderly beta version prevail.

We hope to stimulate informetric researchers with this poster to evaluate and use the Google APIs.

**Acknowledgments**